\documentclass[useAMS,usenatbib,twocolumn]{mn2e}
\usepackage{natbib}
\usepackage{graphicx}
\usepackage{times}
\usepackage{rotate}
\usepackage{color,url}
\newcommand{\simgt}{\lower.5ex\hbox{$\; \buildrel > \over \sim \;$}}
\newcommand{\simlt}{\lower.5ex\hbox{$\; \buildrel < \over \sim \;$}}
\newcommand{\ave}[1]{\left\langle #1\right\rangle}

\begin{document}
\title[Fingers-of-God effect of infalling satellite galaxies]
{Fingers-of-God effect of infalling satellite galaxies}

\author[Hikage]{Chiaki Hikage$^{1,2}$, Kazuhiro Yamamoto $^3$ \\
$^1$ Kavli Institute for the Physics and Mathematics of the Universe (Kavli IPMU, WPI), University of Tokyo, 5-1-5 Kashiwanoha, Kashiwa, Chiba, 277-8583, Japan\\
$^2$ Kobayashi-Maskawa Institute for the Origin of Particles and the Universe (KMI), Nagoya University, 464-8602, Japan\\
$^3$ Department of Physical Sciences, Hiroshima University, Higashi-hiroshima, Kagamiyama 1-3-1, 739-8526, Japan\\
}
\maketitle

\label{firstpage}

\begin{abstract}
Nonlinear redshift-space distortion known as the Fingers-of-God (FoG)
effect is a major systematic uncertainty in redshift-space
distortion studies conducted to test gravity models. The FoG effect has been usually
attributed to the random motion of galaxies inside their
clusters. When the internal galaxy motion is not well virialized,
however, the coherent infalling motion toward the cluster center
generates the FoG effect. Here we derive an analytical model of the
satellite velocity distribution due to the infall motion combined with
the random motion. We show that the velocity distribution becomes far
from Maxwellian when the infalling motion is dominant. We use
simulated subhalo catalogs to find that the contribution of infall
motion is important to massive subhalos and that the velocity
distribution has a top-hat like shape as expected from our analytic
model. We also study the FoG effect due to infall motion on the
redshift-space power spectrum. Using simulated mock samples of
luminous red galaxies constructed from halos and massive subhalos in
N-body simulations, we show that the redshift-space power spectra can
differ from expectations when the infall
motion is ignored.
\end{abstract}


\section{Introduction}
\label{sec:intro}
The peculiar motion of galaxies imprinted on redshift-space clustering
provides a good probe of the dynamics of galaxies. The coherent motion
of galaxies associated with gravitational evolution squashes the
spatial distribution of galaxies along the line-of-sight direction, which is known as the Kaiser
effect \citep{Kaiser87,Hamilton92} and provides a unique probe of
the growth rate \citep{Peacock01}. The growth rate has been
measured from the results of a variety of galaxy surveys to test general relativity
and modified gravity \citep[e.g.,][]{Yamamoto08,Guzzo08,VIPERS,Beutler14,Reid14}.   One can
expect further precise measurement of the growth rate in a wide
range of redshifts from the results of future cosmological surveys such as Subaru/PFS
\citep{PFS}, Euclid \citep{Euclid}, DESI \citep{DESI} and WFIRST
\citep{WFIRST}.

The internal motions of galaxies in their host halos elongate the
distribution of galaxies along the line-of-sight direction, which is
called the Fingers-of-God (FoG) effect \citep{Jackson72}. The FoG effect
depends on the type of galaxy \citep{Zehavi05}. For red galaxy
samples, such as luminous red galaxies (LRGs), the FoG effect is influential
at scales larger than 10$h^{-1}$ Mpc and can introduce serious systematics
in the measurement of the cosmic growth rate
\citep[e.g.,][]{HY13,Beutler14,Reid14}. Clustering anisotropy
can generally be expanded with a series of multipole power spectra
$P_l(k)$ \citep{Yamamoto06}. The Kaiser effect mainly generates quadrupole
anisotropy \citep{Kaiser87}. Meanwhile, the FoG effect generates
higher multipole anisotropy, such as hexadecapole ($l=4$) and
tetra-hexadecapole ($l=6$) components, where the Kaiser effect is
subdominant \citep{HY13}. These multipoles are useful in eliminating
the FoG uncertainty and provide kinematic information of
satellite galaxies \citep{Hikage14,Kanemaru15}.

The FoG effect is related to the kinematics of central and satellite
galaxies inside their host halos. The satellite dynamics can be
altered by different physical processes such as dynamical friction
\citep{Chandrasekhar43}, tidal stripping/disruption
\citep[e.g.,][]{BoylanKolchin08,WetzelWhite10}, satellite merging,
and hydrodynamical drag, such as ram pressure
\citep{GunnGott72}. Such complicated processes of satellite kinematics
introduce systematic uncertainties in the determination of the dynamical mass of
galaxy clusters through satellite velocity dispersion
\citep[e.g.,][]{Wu13}. The velocity distribution of satellites has been
studied by conducting various numerical simulations
\citep[e.g.,][]{Ghigna00,Diemand04,Faltenbacher05,Wu13}. A non-zero
velocity of central galaxies relative to the host halo has been
found in several studies \citep[e.g.,][]{vandenBosch05,Guo15}.

%%%%%%%%%%%%%%%%%%%%%%%%%%%%%%%%%%%%%%%%%%%%%%%%%%%%%%%%
\begin{figure*}
\begin{center}
\includegraphics[width=4cm]{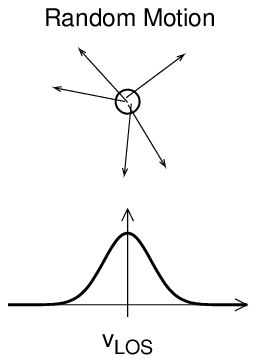}
\includegraphics[width=4cm]{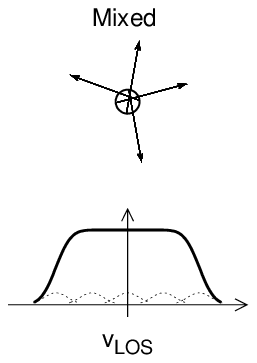}
\includegraphics[width=4cm]{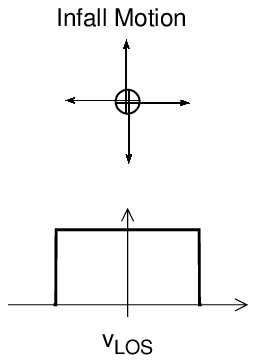}
\caption{Illustration of the line-of-sight velocity distribution $v_{\rm
    LOS}$ due to random motion (left), coherent infall motion
  (right) and their combination (center). The velocity distribution due to
  coherent infall motion has a top-hat shape, while the random
  motion produces a Gaussian velocity distribution.}
\label{fig:vdist}
\end{center}
\end{figure*}
%%%%%%%%%%%%%%%%%%%%%%%%%%%%%%%%%%%%%%%%%%%%%%%%%%%%%%%%

The FoG effect has usually been attributed to the random motion of
galaxies inside clusters or halos. Meanwhile, coherent
infall motion onto the halo mass center also generates the FoG effect,
as illustrated by \cite{Hamilton92}. The infall region of clusters forming a
trumpet-shaped pattern has been observed for a number of clusters and
provides an important probe of the cluster mass profile insensitive to the
details of galaxy formation
\citep{RegosGeller89,DiaferioGeller97,Rines03,ZuWeinberg13}.  In this
letter, we study the FoG effect due to the infall motion of satellites
on the redshift-space power spectrum. We derive a theoretical model of
the line-of-sight velocity distribution due to the random and infall
motions of satellites inside their halos.  We use simulated subhalo
catalogs to test our model expectations and to study the effect on the
redshift-space power spectra. Here we focus on LRGs that have been
widely used in cosmological studies
\citep[e.g.,][]{Reid10,Beutler14,Reid14}. Although the satellite
fraction of LRGs is only about 6\%, the effects of the satellite FoG
effect on the growth rate measurement have been found to be large
\citep{HY13}.  Understanding the behavior of satellite motion is
important in precise cosmological studies using redshift-space
power spectra of LRG samples.

This letter is organized as follows. In section \ref{sec:2}, we model
satellite internal motion by decomposing the infall and randomly
rotating components. The details of our simulations are explained in section
\ref{sec:3}. In section \ref{sec:4}, we present results, comparing the
satellite velocity distribution with the theoretical modeling derived
in section \ref{sec:2}, and evaluate the FoG effects of the infall
velocity and random velocity on redshift-space galaxy
clustering. Section \ref{sec:5} summarizes the letter and presents 
conclusions.

\vspace{-0.5cm}
\section{Satellite motion inside host halos}
\label{sec:2}
The internal satellite velocity with respect to the host halo bulk
velocity ${\mathbf v}_{\rm sat}\equiv {\mathbf V}_{\rm sat}-{\mathbf
  V}_{\rm halo}$ can be decomposed into a component infalling onto the
halo center and a tangential component:
%%%%%%%%%%%%%%%%%%%%%%%%%%%%%%%%%%%%%%%%%%%%%%%%%%%%%%%%
\begin{equation}
{\mathbf v}_{\rm sat} = (-\ave{v_{\rm inf}} + \epsilon_{\rm inf}){\mathbf e}_r 
+ \epsilon_{\rm tan,\theta}{\mathbf e}_\theta + \epsilon_{\rm tan,\phi}{\mathbf e}_\phi,
\end{equation}
%%%%%%%%%%%%%%%%%%%%%%%%%%%%%%%%%%%%%%%%%%%%%%%%%%%%%%%%
where the mean infall velocity $\ave{v_{\rm inf}}$ is generally
non-zero and defined to have a positive sign in the direction toward
the halo center. The average velocity dispersions in the infall and
tangential directions are defined as $\sigma_{\rm v,inf}^2 \equiv
\ave{\epsilon_{\rm inf}^2}$ and $\sigma_{\rm v,tan}^2 \equiv
\ave{\epsilon_{\rm tan,\theta}^2}=\ave{\epsilon_{\rm tan,\phi}^2}$
respectively, and depend on the host halo mass $M$.  The
line-of-sight (one-dimensional) component of the internal satellite velocity
$v_{\rm LOS}\equiv {\mathbf v}_{\rm sat}\cdot{\mathbf e}_{\rm LOS}$
becomes
%%%%%%%%%%%%%%%%%%%%%%%%%%%%%%%%%%%%%%%%%%%%%%%%%%%%%%%%
\begin{equation}
v_{\rm LOS}=(-\ave{v_{\rm inf}}+\epsilon_{\rm inf})\mu+\epsilon_{\rm tan,\theta}(1-\mu^2)^{1/2},
\end{equation}
%%%%%%%%%%%%%%%%%%%%%%%%%%%%%%%%%%%%%%%%%%%%%%%%%%%%%%%%
where $\mu$ is the cosine of the angle between the line of sight
and the direction of infall toward the halo center; i.e.,
$\mu\equiv{\mathbf e}_r\cdot {\mathbf e}_{\rm LOS}$.  When the infall
and tangential velocities are given by independent Gaussian distributions, the
line-of-sight velocity distribution of satellites in the direction of $\mu$
becomes a Gaussian distribution with a mean of $-\mu\ave{v_{\rm inf}}$ and 
variance of
%%%%%%%%%%%%%%%%%%%%%%%%%%%%%%%%%%%%%%%%%%%%%%%%%%%%%%%%
\begin{equation}
\sigma^2_{\rm v,\mu}(\mu;M)= \mu^2 \sigma^2_{\rm v,inf}(M) + 
(1-\mu^2)\sigma^2_{\rm v,tan}(M).
\end{equation}
%%%%%%%%%%%%%%%%%%%%%%%%%%%%%%%%%%%%%%%%%%%%%%%%%%%%%%%%
The line-of-sight velocity distribution averaged over $\mu$ is given by
%%%%%%%%%%%%%%%%%%%%%%%%%%%%%%%%%%%%%%%%%%%%%%%%%%%%%%%%
\begin{eqnarray}
\label{eq:fv_sat}
f_{\rm v}(v_{\rm LOS};M)&=&\frac{1}{2}\int_{-1}^{1} d\mu
\frac{1}{(2\pi)^{1/2}\sigma_{\rm v,\mu}(\mu;M)} \nonumber \\
&& \times \exp\left[-\frac{(v_{\rm LOS}+\mu\ave{v_{\rm inf}}(M))^2}{2\sigma_{\rm
      v,\mu}^2(\mu;M)}\right].
\end{eqnarray}
%%%%%%%%%%%%%%%%%%%%%%%%%%%%%%%%%%%%%%%%%%%%%%%%%%%%%%%%
The velocity dispersion of satellites is the second moment
of the equation (\ref{eq:fv_sat}) and becomes
%%%%%%%%%%%%%%%%%%%%%%%%%%%%%%%%%%%%%%%%%%%%%%%%%%%%%%%%
\begin{equation}
\sigma_{\rm v,LOS}^2(M)=\frac{1}{3}(\langle v_{\rm inf} \rangle^2+\sigma_{\rm v,inf}^2+2\sigma_{\rm v,tan}^2).
\end{equation}
%%%%%%%%%%%%%%%%%%%%%%%%%%%%%%%%%%%%%%%%%%%%%%%%%%%%%%%%

When the internal motion of galaxies is well virialized, the mean
infall velocity is negligible compared with the random motion and the
internal velocity distribution is isotropic, i.e., $\sigma_{\rm
  v,inf}=\sigma_{\rm v,tan}=\sigma_{\rm vir}$, where $\sigma_{\rm vir}$
is the virial velocity dispersion. Equation (\ref{eq:fv_sat})
simply becomes the Gaussian distribution
%%%%%%%%%%%%%%%%%%%%%%%%%%%%%%%%%%%%%%%%%%%%%%%%%%%%%%%%
\begin{equation}
\label{eq:fv_ga}
f_v(v_{\rm LOS};M)=\frac{1}{(2\pi)^{1/2}\sigma_{\rm v,vir}^2}
\exp\left[-\frac{v_{\rm vir}^2}{2\sigma_{\rm v,vir}^2}\right],
\end{equation}
%%%%%%%%%%%%%%%%%%%%%%%%%%%%%%%%%%%%%%%%%%%%%%%%%%%%%%%%
which is illustrated in the left panel of Figure \ref{fig:vdist}.
Meanwhile, when the mean infall velocity is dominant ($\ave{v_{\rm
  inf}}\gg \sigma_{\rm v,inf},\sigma_{\rm v,tan}$), equation
(\ref{eq:fv_sat}) becomes a top-hat distribution:
%%%%%%%%%%%%%%%%%%%%%%%%%%%%%%%%%%%%%%%%%%%%%%%%%%%%%%%%
\begin{equation}
f_v(v_{\rm LOS};M)=\left\{
\begin{array}{ll}
0.5\langle v_{\rm inf}\rangle^{-1} & (|v_{\rm LOS}|\le\langle v_{\rm inf}\rangle) \\
0 & (|v_{\rm LOS}|>\langle v_{\rm inf}\rangle) 
\end{array}
\right.,
\end{equation}
%%%%%%%%%%%%%%%%%%%%%%%%%%%%%%%%%%%%%%%%%%%%%%%%%%%%%%%%
where the line-of-sight velocity dispersion becomes $\sigma_{\rm
  v,LOS}=\langle v_{\rm inf}\rangle/\sqrt{3}$ (see the right panel in
Figure \ref{fig:vdist}). In reality, the infall motion and random
motion combine and the line-of-sight velocity distribution is
then described by the sum of Gaussian distributions with different mean values of
$\mu\ave{v_{\rm inf}}$ in the range of $\mu$ from $-1$ to $1$. The
shape of the distribution then becomes a smoothed top-hat function like that
seen in the center panel of Figure \ref{fig:vdist}.

%%%%%%%%%%%%%%%%%%%%%%%%%%%%%%%%%%%%%%%%%%%%%%%%%%%%%%%%
\begin{figure*}
\begin{center}
\includegraphics[width=5.3cm]{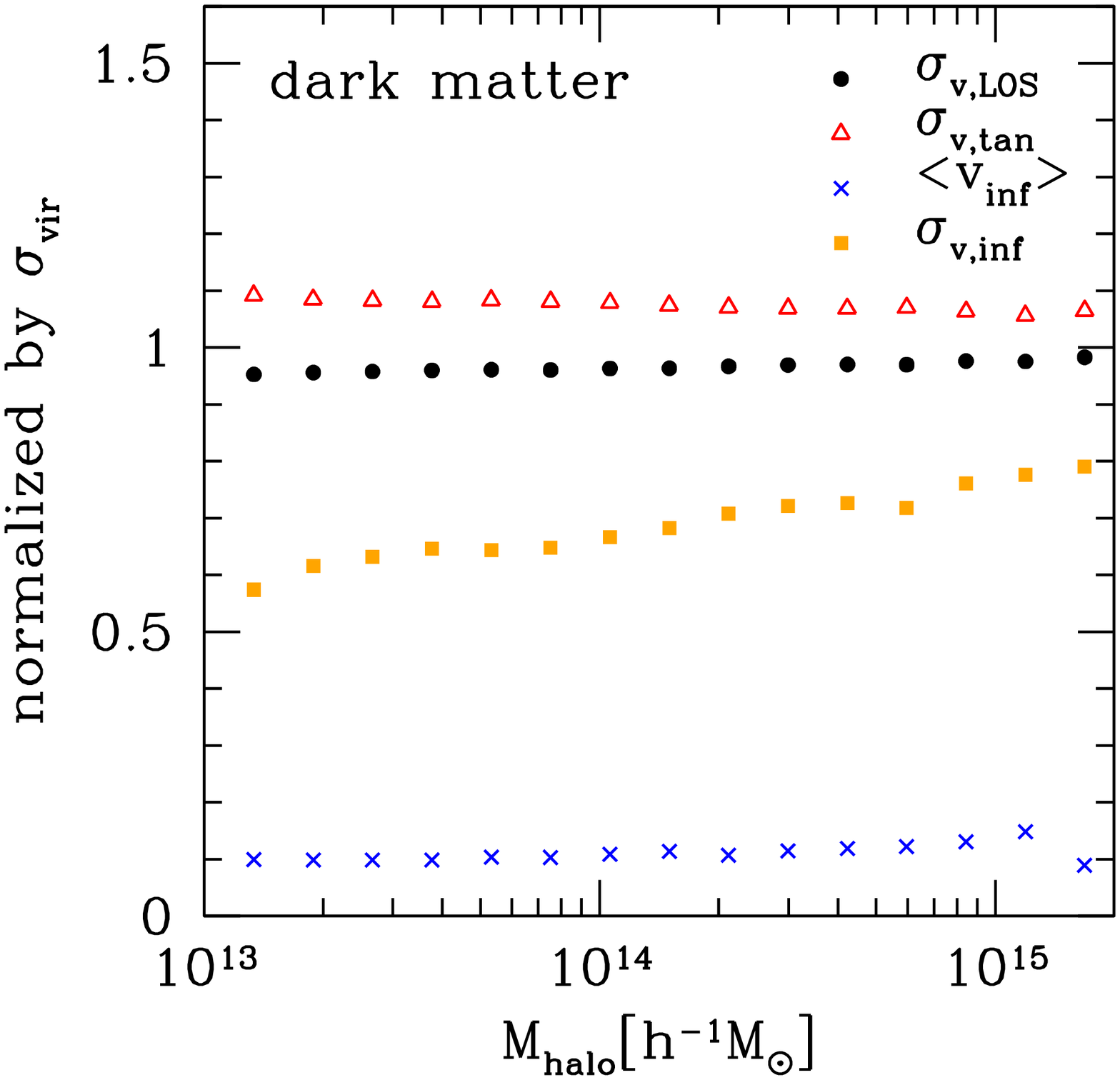}
\includegraphics[width=5.3cm]{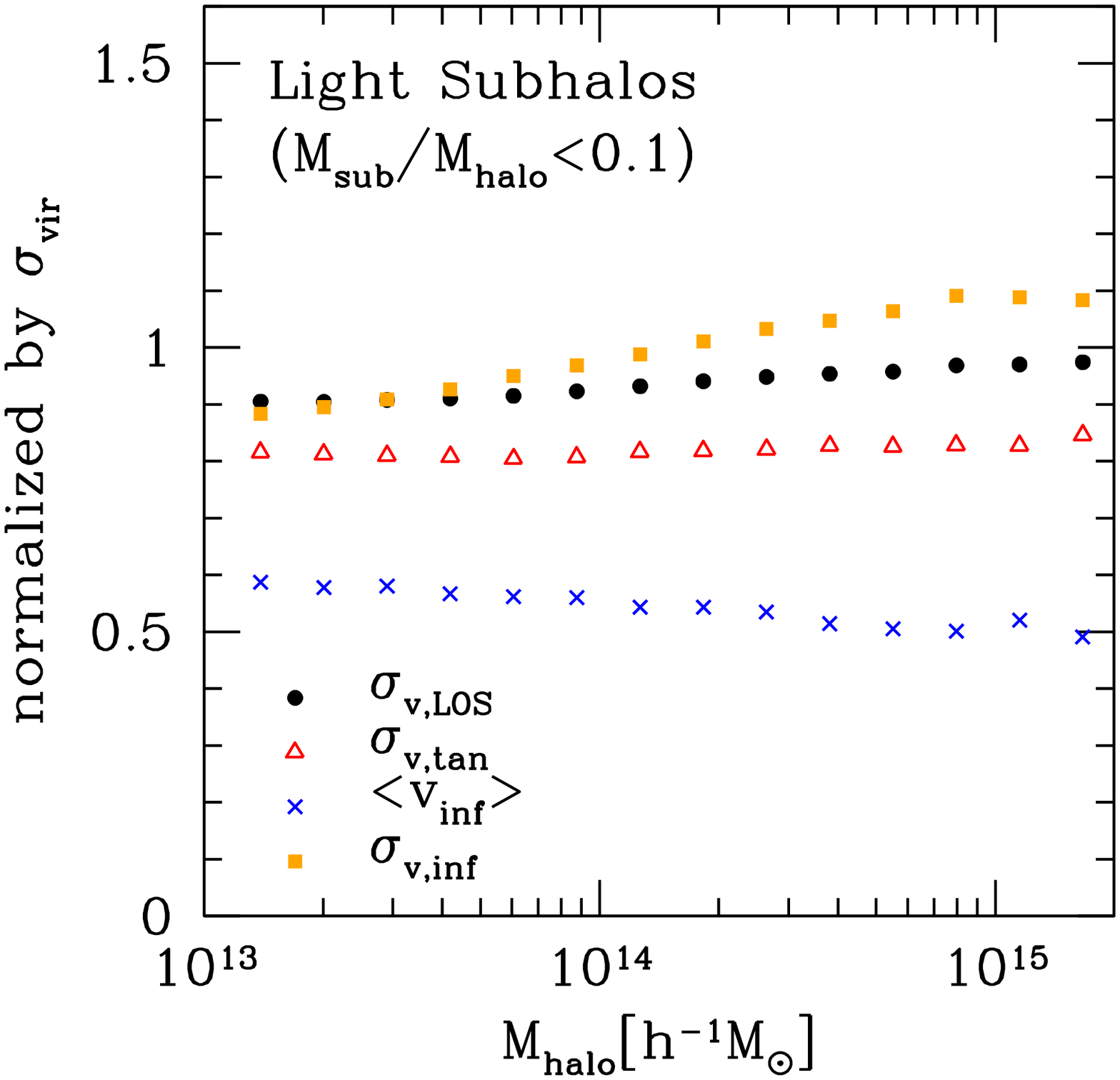}
\includegraphics[width=5.3cm]{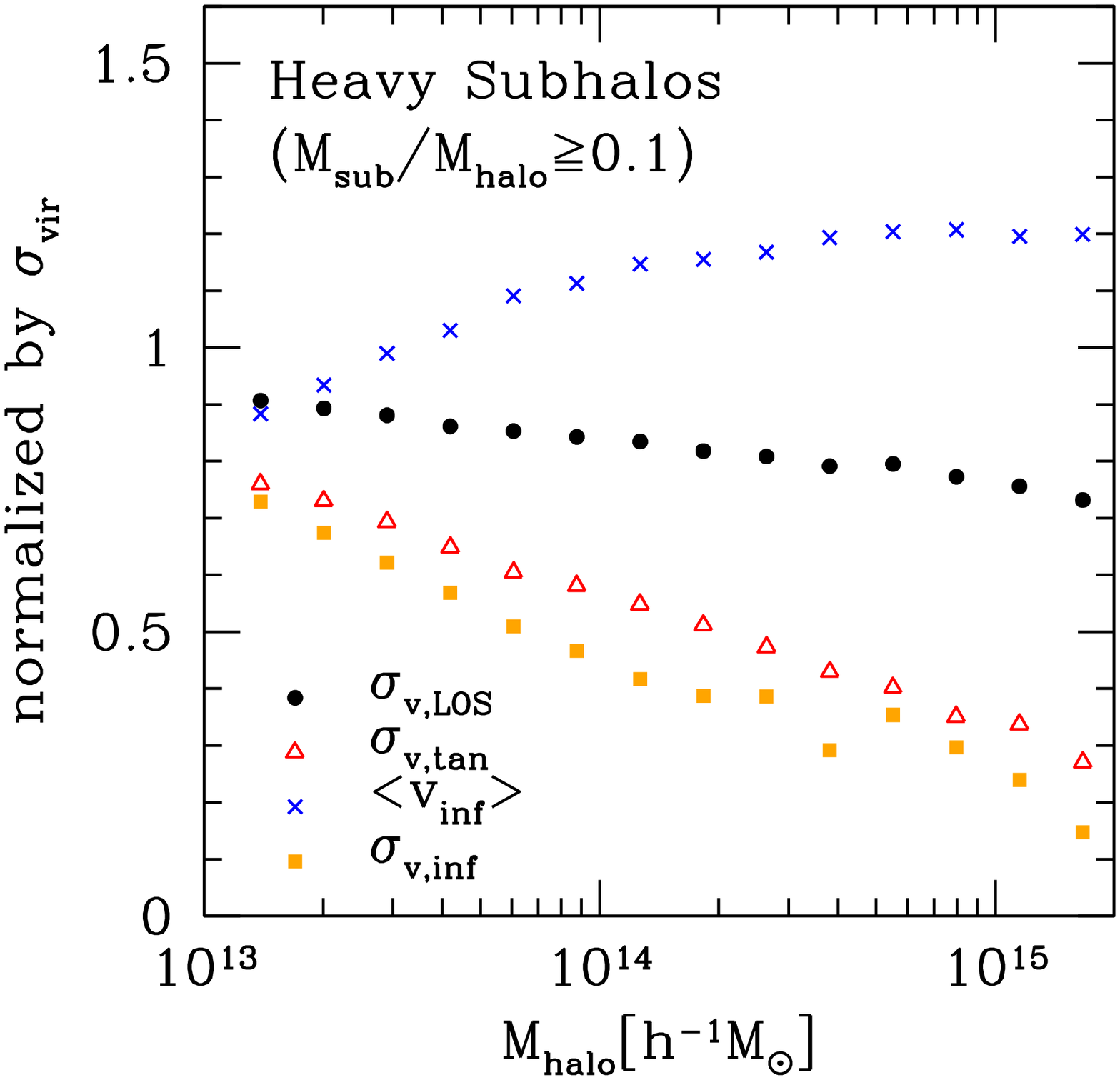}
\caption{Halo mass dependence of the line-of-sight (one-dimensional) velocity
  dispersion $\sigma_{\rm v,LOS}$ (black circles), radial and
  tangential velocity dispersions $\sigma_{\rm v,inf}$ (yellow square)
  and $\sigma_{\rm v,tan}$ (red triangle) and mean infall velocity
  $\ave{v_{\rm inf}}$ (blue crosses) relative to the virial velocity
  dispersion $\sigma_{\rm vir}$.}
\label{fig:sigv}
\end{center}
\end{figure*}
%%%%%%%%%%%%%%%%%%%%%%%%%%%%%%%%%%%%%%%%%%%%%%%%%%%%%%%%

\vspace{-0.5cm}
\section{Simulations}
\label{sec:3}
To validate our modeling of the line-of-sight velocity
distribution derived in the previous section, we construct subhalo
catalogs from N-body simulations. We make 10 realizations using
GADGET-2 public code \citep{Springel05}, starting from an initial
redshift of 80 and a 2LPT initial condition \citep{Crocce06}. The side
length of the simulation box is 600$h^{-1}$ Mpc and the number of
particles is $800^3$, which corresponds to a particle mass of
$2.8\times 10^{10}h^{-1} M_\odot$.  The softening length is set to 
10$h^{-1}$ kpc.  We compute the initial power spectrum using CAMB
software \citep{Lewis00} by assuming a flat $\Lambda$CDM cosmology
with the parameters $\Omega_m=0.273$, $\Omega_b=0.046$,
$h=0.704$, $n_s=0.963$, $\tau=0.089$, and $\sigma_8=0.809$. We use 
snapshot data at $z=0.3$ roughly corresponding to the mean redshift of
the LRG sample of the Sloan Digital Sky Survey (SDSS)
\citep{Eisenstein01}. Halos are identified using the friends-of-friends
algorithm with a linking length $b=0.2$ and setting the minimum number of
particles for halos to 32.  Subhalos are identified using the
SUBFIND code \citep{Springel01} with the minimum number of particles
set to 20. We take the $z$-axis as the line-of-sight direction to add
the peculiar velocity effect as $r\rightarrow r+(1+z)v_z/aH(z)$.

We describe the relationship between halos and galaxies using
the halo occupation distribution (HOD)
\citep{Kravtsov04,Zehavi05,White11}.  We use a
conventional HOD form with five parameters \citep{Zheng05}:
%%%%%%%%%%%%%%%%%%%%%%%%%%%%%%%%%%%%%%%%%%%%%%%%%%%%%%%%
\begin{eqnarray}
\langle N_{\rm cen}(M)\rangle &=&\frac{1}{2}\left[1+{\rm erf}\left(\frac{\log_{10}(M)-\log_{10}
(M_{\rm min})}{\sigma_{\log M}}\right)\right], \\
\langle N_{\rm sat}(M)\rangle &=&\langle N_{\rm cen}\rangle \left(\frac{M-M_{\rm cut}}{M_1}\right)^{\alpha},
\label{eq:HOD}
\end{eqnarray}
%%%%%%%%%%%%%%%%%%%%%%%%%%%%%%%%%%%%%%%%%%%%%%%%%%%%%%%%
where ${\rm erf}(x)$ is the error function. We fix the HOD values for an
SDSS LRG as \citep{Reid09a} $M_{\rm min}=5.7\times
10^{13}h^{-1} M_\odot$, $\sigma_{\log M}=0.7$, $M_{\rm cut}=3.5\times
10^{13}h^{-1} M_\odot$, $M_{\rm 1}=3.5\times 10^{14}h^{-1} M_\odot$, and
$\alpha=1$. The position of the central galaxy is assigned to be the
potential minimum of the halo and the velocity is assigned as the mean
of all dark matter particles inside central subhalos. We use three
different types of tracers for satellite galaxies; one type is randomly
selected dark matter particles and the others are subhalos with
mass relative to the host halo mass $f_{\rm sub}\equiv M_{\rm
  sub}/M_{\rm halo}$ of $<$0.1 and $\ge 0.1$ (hereafter referred to 
as ``light subhalos'' and ``heavy subhalos'' respectively).

\vspace{-0.75cm}
\section{Results}
\label{sec:4}

%%%%%%%%%%%%%%%%%%%%%%%%%%%%%%%%%%%%%%%%%%%%%%%%%%%%%%%
\begin{figure*}
\begin{center}
\includegraphics[width=5.3cm]{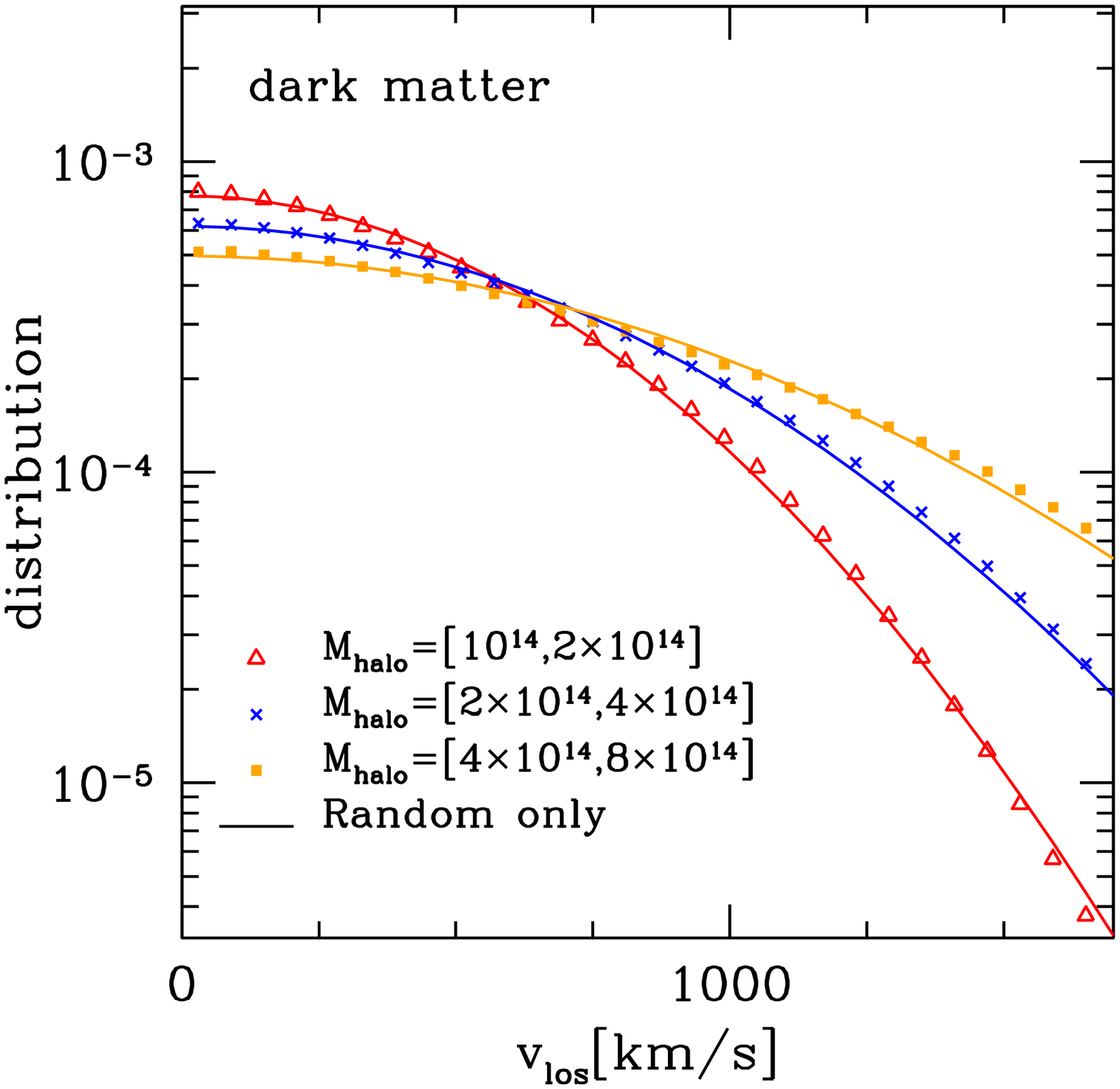}
\includegraphics[width=5.3cm]{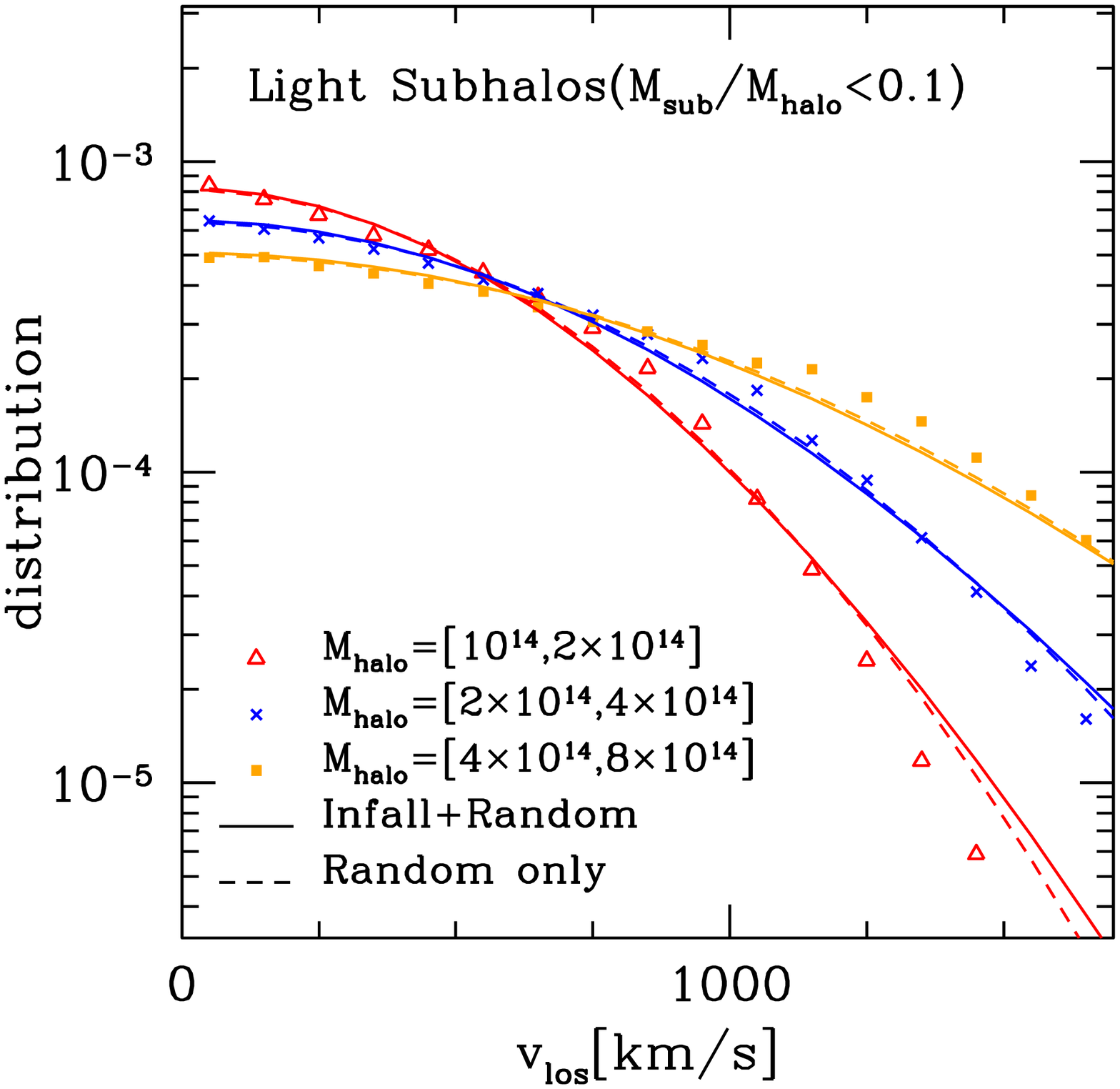}
\includegraphics[width=5.3cm]{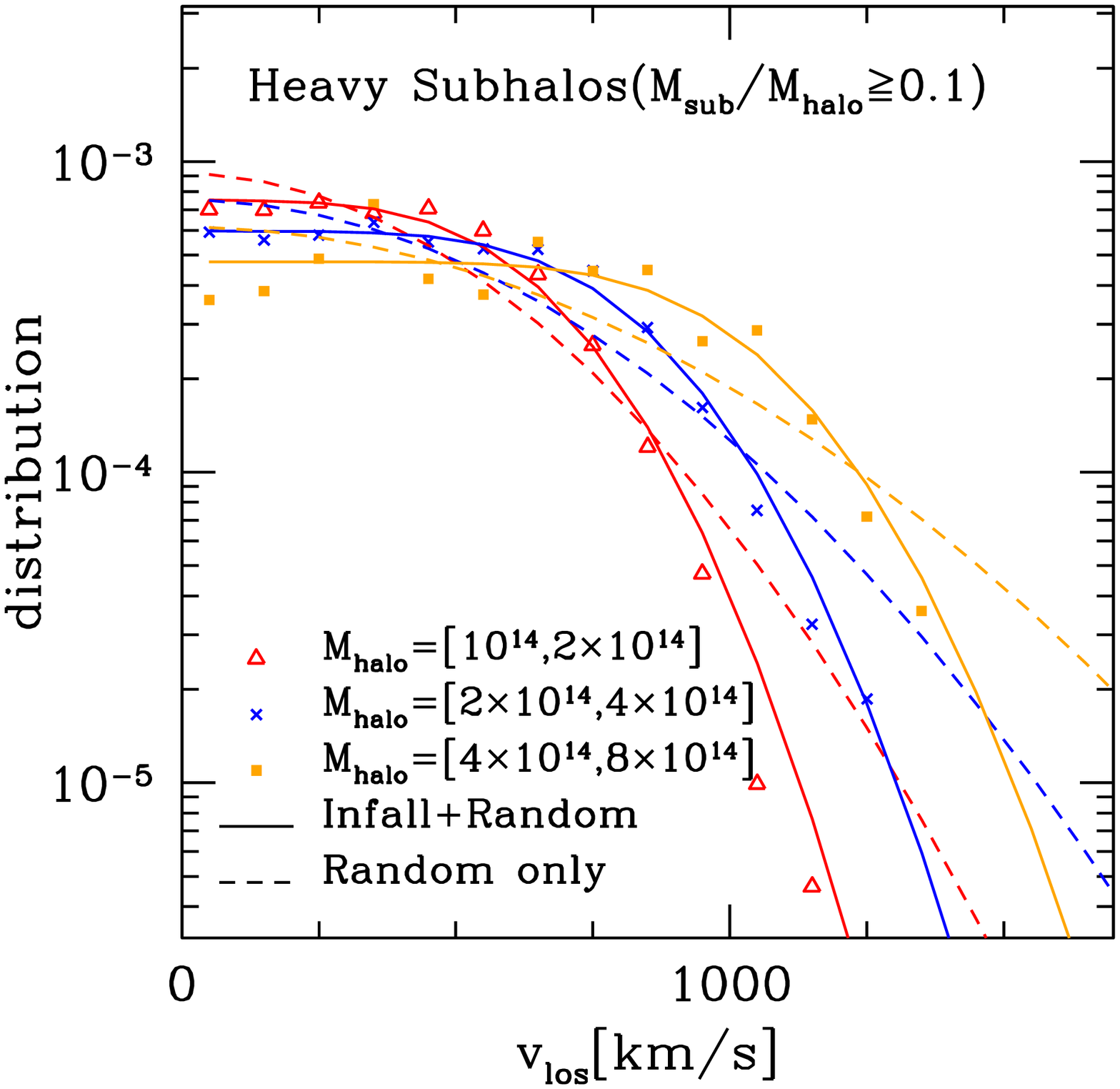}
\caption{Distribution function of the line-of-sight velocity $v_{\rm
    los}$ distribution of satellites for three different ranges of
  halo mass from $10^{14}h^{-1} M_\odot$ to $8\times
  10^{14}h^{-1} M_\odot$.  Satellites are represented with dark matter particles
  (left), light subhalos (center) and heavy subhalos (right).
  Reference lines are a Gaussian distribution of the one-dimensional velocity
  dispersion (eq. [\ref{eq:fv_ga}], dashed) and our velocity model
  including both the infall and random motions
  (eq. [\ref{eq:fv_sat}], solid).}
\label{fig:veldist}
\end{center}
\end{figure*}
%%%%%%%%%%%%%%%%%%%%%%%%%%%%%%%%%%%%%%%%%%%%%%%%%%%%%%%%

\subsection{Satellite velocity distribution}
We use simulated subhalo catalogs and decompose the internal velocity
of satellites relative to the host halo velocity into infall and
tangential velocity components. Figure \ref{fig:sigv} shows the results
of the host halo mass dependence of each velocity component:
the line-of-sight velocity dispersion $\sigma_{\rm v,LOS}$, mean infall
velocity $\ave{v_{\rm inf}}$ and the infall and tangential velocity
dispersion $\sigma_{\rm v,inf}$ and $\sigma_{\rm v,tan}$. The panels
show the result when satellites are represented using dark matter
particles, light subhalos, and heavy subhalos from left to right.
Each velocity component is normalized with the virial velocity
dispersion $\sigma_{\rm vir}$ for the corresponding host halo mass.

When satellites are represented by dark matter particles, the mean
infall velocity is low and the line-of-sight velocity dispersion is
nearly equal to the virial velocity dispersion. This indicates that
the motion of dark matter particles is well virialized. Meanwhile, the
motion of subhalos is quite different from that of dark matter
particles. The mean infall velocity of subhalos is much higher than that of 
the dark matter particles and it is comparable to the virial velocity for
massive subhalos. The tangential velocity reduces as the
mean infall velocity increases. This indicates that the dynamical
friction becomes efficient for heavier subhalos and the orbital motion
slows down. The velocity bias of the line-of-sight velocity dispersion
varies depending on the host halo mass and the subhalo mass. Our result
shows that the line-of-sight velocity dispersion is smaller than the
virial velocity dispersion. The result may change quantitatively when
the subhalo finders are different.  The inner fast rotating subhalos
are difficult to identify using the SUBFIND algorithm and the
average velocity of satellite subhalos is thus lower than that in 
previous works \citep[e.g.,][]{Diemand04}. The qualitative result that
the infall motion is dominant in heavier subhalos is consistent with
the results obtained previously.

Figure \ref{fig:veldist} shows the line-of-sight velocity distribution
of satellites represented by dark matter particles (left), light
subhalos (center) and heavy subhalos (right) for different bins of
host halo mass. The velocity distribution for dark matter satellites
agrees with a Gaussian distribution with virial velocity dispersion
for each bin of the halo mass. For satellite subhalos, the mean infall
velocity is comparable to or larger than the tangential velocity. The
resulting shape of the line-of-sight velocity distribution does not
become a simple Gaussian but has a more top-hat-like shape,
particularly for heavy subhalos. This feature is consistent with
previous results obtained by \cite{Diemand04}, who pointed out that
the subhalo velocity distribution becomes non-Maxwellian. We compare
the simulation results with our modeling using the simulated values of
the mean infall velocity and velocity dispersions.  For heavy subhalos
(the right panel of Figure \ref{fig:veldist}), our modeling of the
line-of-sight velocity distribution given by equation
(\ref{eq:fv_sat}) describes the simulation results better than a
simple Gaussian distribution (eq. [\ref{eq:fv_ga}]).

\subsection{FoG effect on the redshift-space power spectrum}
%%%%%%%%%%%%%%%%%%%%%%%%%%%%%%%%%%%%%%%%%%%%%%%%%%%%%%%%
\begin{figure}
\begin{center}
\includegraphics[width=6cm]{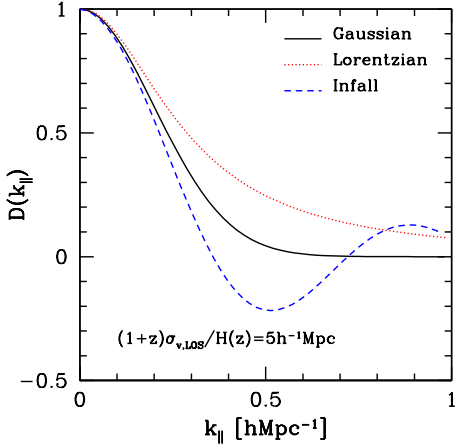}
\caption{Comparison of the FoG damping function ${\cal
    D}(k_\parallel)$ (eq. [\ref{eq:fogdamp}]) when the infall motion or
  random motion is dominant. For comparison, we plot the
  Lorentzian form of FoG damping.  The velocity dispersion
  $\sigma_{\rm v,LOS}/aH(a)$ is set to 5$h^{-1}$ Mpc.}
\label{fig:fogfunc}
\end{center}
\end{figure}
%%%%%%%%%%%%%%%%%%%%%%%%%%%%%%%%%%%%%%%%%%%%%%%%%%%%%%%%

%%%%%%%%%%%%%%%%%%%%%%%%%%%%%%%%%%%%%%%%%%%%%%%%%%%%%%%%
\begin{figure}
\begin{center}
\includegraphics[width=6cm]{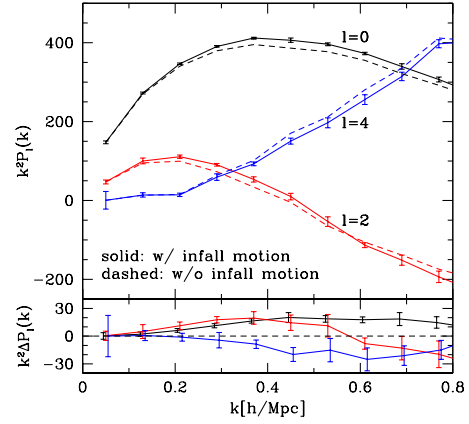}
\caption{Mock LRG power spectra $P_l$ simulated with the halo model
  description. Satellites are represented by massive subhalos (where the average
  ratio of the subhalo mass to the host halo mass is 0.06) and the
  satellite fraction is about 6.4\%. For comparison, we plot the power
  spectra where the velocity of satellite subhalos is rotated randomly.}
\label{fig:pkl}
\end{center}
\end{figure}
%%%%%%%%%%%%%%%%%%%%%%%%%%%%%%%%%%%%%%%%%%%%%%%%%%%%%%%%

Satellite motion inside halos generates non-linear redshift-space
distortion, often referred as the FoG effect. The FoG
effect on multipole components of the redshift-space power spectrum is
formulated employing the halo model \citep[e.g.,][]{Hikage13,HY13}. 
FoG damping due to internal satellite motion of the halo with mass
$M$ is given by Fourier transforming the line-of-sight velocity
distribution of satellites:
%%%%%%%%%%%%%%%%%%%%%%%%%%%%%%%%%%%%%%%%%%%%%%%%%%%%%%%%
\begin{eqnarray}
\label{eq:fogdamp}
{\cal D}(k_\parallel;M)=
\int dv f_v(v;M)\exp(-i\tilde{k}_\parallel v)~~~~~~~~~~~ \nonumber \\
~~~~~~~~\simeq \exp\left[-\frac{\tilde{k}_\parallel^2 (\sigma_{\rm v,sat}^2-\ave{v_{\rm inf}}^2/3)}{2}\right]
\frac{\sin (\tilde{k}_\parallel\ave{v_{\rm inf}})}{\tilde{k}_\parallel\ave{v_{\rm inf}}},
\end{eqnarray}
%%%%%%%%%%%%%%%%%%%%%%%%%%%%%%%%%%%%%%%%%%%%%%%%%%%%%%%%
where $\tilde{k}=(1+z)k_\parallel/H(z)$ and $k_\parallel$ is the
line-of-sight component of the wavevector $\mathbf{k}$. In the second
line, we make the approximation that $\sigma_{\rm v,inf}\simeq
\sigma_{\rm v,tan}$. Figure \ref{fig:fogfunc} compares the FoG damping
function in different cases. When the mean infall velocity is
equal to zero, FoG damping has a simple Gaussian form.
When the infall velocity is dominant, the velocity distribution has a
top-hat form and its Fourier-transform is a sinc function, as shown
by the dashed line in Figure \ref{fig:fogfunc}. FoG damping due to
the infall motion has a shape quite different from the commonly used
Gaussian or Lorentzian form and affects the small-scale features of
redshift-space clustering.

Figure \ref{fig:pkl} shows the multipole power spectra $k^2P_l(k)$
($l=0,2,4$ from left to right) of simulated samples assuming the HOD
of SDSS LRGs. Satellites in the samples are represented by heavy
subhalos, which is a good assumption because LRGs are massive galaxies
and thus should have massive host subhalos. The average ratio of
the subhalo mass to the host halo mass becomes 6.4\%. In this sample,
the mean infall velocity is comparable to the virial velocity of
satellite-hosting halos and the values of each velocity component of
satellite subhalos are $\langle v_{\rm
  inf}\rangle=0.79\sigma_{\rm vir}$; $\sigma_{\rm
  v,inf}=0.95\sigma_{\rm vir}$; $\sigma_{\rm v,tan}=0.86\sigma_{\rm
  vir}$. To see the effect of coherent infall motion, we also
plot the power spectra of the same mock samples where the direction of
the internal satellite velocity $\mathbf{v}_{\rm sat}$ is randomly
rotated to erase the coherent infall motion while the absolute value
$|\mathbf{v}_{\rm sat}|$ is the same. The error bars represent the 1-sigma
scatter of the power spectra $P_l$ for 10 LRG mock samples divided by
the square root of 10.  Assuming that the sample variance is
proportional to the survey volume, the plotted error corresponds to
the statistical error for the survey with the sample volume of
2.16$h^{-3} $Gpc$^3$ ($=10\times (600h^{-1}{\rm Mpc})^3$), which is
roughly the same volume as for the BOSS/CMASS sample \citep{Beutler14}. The effect
of the infalling motion of satellites becomes appreciable on smaller
scales ($k\simgt 0.2h {\rm Mpc}^{-1}$) because the internal velocity distribution
is sensitive to the small-scale power spectrum via the FoG
effect. We evaluate the effect of the infall FoG on the overall shape of
multipole power spectra as chi-square values:
%%%%%%%%%%%%%%%%%%%%
\begin{equation}
\chi^2 \equiv \sum_l^{0,2,4}\sum_i^{k_i<k_{\rm max}} 
\frac{[P_l^{\rm w/ infall}(k_i)-P_l^{\rm w/o~infall}(k_i)]^2}{\sigma_{P_l}^2(k_i)},
\end{equation}
%%%%%%%%%%%%%%%%%%%%
where $\sigma_{P_l}$ is the 1$\sigma$ error of $P_l$ for the survey with 
$2.16h^{-3}$ Gpc$^3$ volume, and $k_{\rm
  max}$ is the maximum value of $k$ for the $\chi^2$ calculation.  The
binning width of $k$ for the $\chi^2$ calculation is set to be
$0.04h {\rm Mpc}^{-1}$. The chi-square value increases at larger $k_{\rm max}$:
$\chi^2=10$ at $k_{\rm max}=0.21h {\rm Mpc}^{-1}$, $\chi^2=89$ at $k_{\rm
  max}=0.29h {\rm Mpc}^{-1}$, and $\chi^2=190$ at $k_{\rm max}<0.41h {\rm Mpc}^{-1}$.  Our
result indicates that the proper treatment of the infall FoG would be
necessary for the precise modeling of redshift-space power spectra
for massive galaxies such as LRGs at $k\simgt 0.2h {\rm Mpc}^{-1}$.

\vspace{-0.5cm}
\section{Summary and Conclusions}
\label{sec:5}
We investigated satellite motion inside host halos and the
resulting FoG effect on the power spectrum. When the
satellite motion is not well virialized, the internal satellite motion
is not random but has coherent infalling flow onto the halo center. We
used subhalo catalogs to find that the infall motion becomes important
when the subhalo mass relative to the host halo mass increases and
the line-of-sight velocity distribution then deviates from Maxwellian
and has a flat top-hat form. We derived a theoretical
model of the satellite velocity including both infall and random motions.
We found that our model well describes the simulated subhalo velocity
distribution. We also investigated how the difference in the velocity
structure between random motion and infall motion affects the shape of
redshift-space power spectra. We showed that the effect of infall
motion on the overall shape of an LRG-like power spectrum is appreciable on the scale $k>0.2h$ per Mpc for the current survey volume, 
such as the BOSS/CMASS sample volume. 

The effect of infall motion may be important in precision
cosmological studies using redshift-space galaxy clustering in future
surveys such as Subaru/PFS, DESI, Euclid, and WFIRST. We found that the
contribution of infall motion depends on the mass of subhalos hosting
satellites. Galaxy--galaxy lensing provides the information of subhalo
mass hosting satellite galaxies \citep[e.g.,][]{Li14}.  It would be
interesting to see how the galaxy--galaxy lensing measurement from the
upcoming imaging and redshift surveys can remove the uncertainty of
the infall motion effect. We leave this as a topic of research to be addressed in the near future.

\vspace{-0.5cm}

\section*{Acknowledgments}
We thank Takahiko Matsubara for useful discussions.
This work was supported by MEXT/JSPS KAKENHI Grant Numbers
24740160 and 15H05895.

\vspace{-0.5cm}

\bibliographystyle{mn2e} \bibliography{mn-jour,ref} \label{lastpage}

\end{document}